# On the gauge dependence of Quantum Electrodynamics

**Henry Kißler**[*]
*Department of Mathematics, Humboldt-Universität zu Berlin, Rudower Chaussee 25, 12489 Berlin, Germany*
*E-mail:* kissler@physik.hu-berlin.de

This paper reports on our diagrammatic approach to characterize the gauge dependence of Quantum Electrodynamics in the linear covariant gauge. Our dimensionally independent technique is purely based on a perturbative analysis and allows us to construct the full expansion in the gauge parameter of a Green's function from its value in a particular gauge (such as the Feynman or Landau gauge). Further, we clarify the compatiblity of perturbation theory and the Landau–Khalatnikov–Fradkin transform.

---

[*]Speaker.

## 1. Introduction

We consider the Lagrangian of Quantum Electrodynamics

$$\mathscr{L} = -\frac{1}{4}F_{\mu\nu}F^{\mu\nu} + \overline{\psi}\left(i\slashed{\partial} - e\slashed{A} - m\right)\psi, \tag{1.1}$$

where $F$ denotes the field-strength tensors of the gauge field, $\psi$ is the fermionic field that interacts through the covariant derivative with the gauge field. It should be remarked that the diagrammatic techniques that will be introduced in the following apply to the massive as well to the massless fermion case. However, in the following explicit evaluations of Feynman graphs, we will always consider the massless case for the sake of simplicity. Perturbation theory requires the introduction of a gauge fixing and due to renormalizability the gauge fixing introduces a gauge parameter that is renormalized in straight analogy to the coupling paramater $\alpha$. In the scope of this article, we will restrict ourselves to the linear covariant gauge

$$\mathscr{L}_{\text{GF}} = -\frac{1}{2\xi}\left(\partial_\mu A^\mu\right)^2, \tag{1.2}$$

and denote the gauge parameter by $\xi$. Such a gauge fixing breaks gauge invariance and therefore off-shell Green's functions are generally gauge dependent objects. In our case, this means that Green's functions depend on the gauge parameter in a non-trivial way. Although, these terms are expected to be non-physical and eventually vanish in the physical on-shell limit they enter perturbative computations and the renormalization process at any stage and potentially limit our ability to perform high loop computations in their full generality.

On the other hand, a non-perturbative investigation of the gauge dependence has been conducted by Landau, Khalatnikov [1], Fradkin [2], and Sonoda [3]. This approach is known as Landau–Khalatnikov–Fradkin transform (LKFT) and allows for a characterization of the gauge dependence of the full Green's function that should be valid to all orders when perturbatively expanded.

The goals of this paper is to clarify how the LKFT is compatible with perturbation theory and Feynman graphs in particular. The key observation is that the sum of all connected Feynman graphs can be partitioned in a way such that various cancellations apply and the gauge dependent contributions eventually only couple to the external legs in the remaining graphs. Conclusively, the gauge dependent terms are characterized by a Dyson–Schwinger type equation (4.2), which matches the LKF formula when transformed into position space.

This purely perturbative characterization of the gauge dependence determines all terms which depend on the gauge parameter in the epsilon expansion of a unrenormalized Green's function. Further, a similar result is obtained for renormalized Green's functions. In the latter case, we will point out that the gauge dependent terms can be factorized into an exponential. This conclusion is one of the main observations in the LKFT approach and has been exploited by Grozin [4], for example, to show that the anomalous dimension of the electron depends on the gauge parameter at the first loop order only. Here, this statement is independently derived by arguments purely based on perturbation theory.

## 2. The Landau–Khalatnikov–Fradkin transform

A non-perturbative investigation of the gauge dependence has been conducted by Landau and Khalatnikov [1] and Fradkin [2] a long time ago. Their discussion can mainly be summarized as follows. Let $S$ be electron propagator in some specific gauge. The gauge of $S$ can be altered by applying a gauge transformation

$$\left.\begin{aligned} A_\mu(x) &\mapsto A_\mu(x) + \partial_\mu \omega(x) \\ \psi(x) &\mapsto e^{-ie\omega(x)} \psi(x) \\ \overline{\psi}(x) &\mapsto \overline{\psi}(x) e^{ie\omega(x)} \end{aligned}\right\} \quad (2.1)$$

to its external legs. Recall that a general gauge transformation is parametrized by a scalar field $\omega$. In the next step, this scalar field is quantized and identified with the longitudinal part of the photon propagator. This establishes the following relation between the propagator $S$ and its gauge-transformed equivalent

$$S'(x-y) = S(x-y) \exp\left[ieM(x-y) - ieM(0)\right]. \quad (2.2)$$

Here, the function $M$ absorbed the dynamics of the scalar field $\omega$ and hence provides a complete characterization of how the electron propagator changes under the considered change of gauge. Closed expressions of the function $M$ for different gauge transitions have been derived by Zumino [5] including the transition from the Landau to the general linear covariant gauge

$$M(x) = \xi \int \frac{d^D p}{(2\pi)^D} \frac{e^{-ip \cdot x}}{(p^2)^2}. \quad (2.3)$$

It should be remarked that the transition from the Feynman gauge to the general linear covariant gauge is similarly described by shifting the gauge parameter in the above expression by $-1$. Now, equation (2.2) implies that the electron propagator in the general covariant gauge can be factorized into its Landau gauge (or Feynman gauge) equivalent and a part that depends on the gauge parameter. Crucially, we observe that the $\xi$ terms can be *factorized into an exponential*.

## 3. A perturbative check of the gauge exponentiation

This section provides a first perturbative check on the factorization and exponentiation of the gauge parameter which is predicted by the Landau-Khalatnikov-Fradkin transform (2.2). For this purpose, we examine the quenched massless self-energy of the electron to fourth order in perturbation theory. We use QGRAF [6], FORM [7, 8], and data of the MINCER [9, 10] and FORCER [11, 12] packages and follow the program of Hopf-algebraic renormalization as described in [13]. $\widetilde{\text{MOM}}$ is utilized as renormalization scheme, that is to say counterterms are constructed in such a way that all quantum corrections to the propagator vanish when its external momentum is evaluated at the renormalization point $-\mu^2$ and the same applies to the vertex if the external photon momentum vanishes.

$$\Sigma(\alpha, \xi, L)/\!\!\!p = \xi L \left(\frac{\alpha}{4\pi}\right) + \left(-\frac{1}{2}\xi^2 L^2 - \frac{3}{2}L\right)\left(\frac{\alpha}{4\pi}\right)^2 + \left(\frac{1}{6}\xi^3 L^3 + \frac{3}{2}\xi L^2 + \frac{3}{2}L\right)\left(\frac{\alpha}{4\pi}\right)^3$$
$$+ \left(-\frac{1}{24}\xi^4 L^4 - \frac{3}{4}\xi^2 L^3 - \left[\frac{3}{2}\xi + \frac{9}{8}\right]L^2 - \left[\frac{1027}{8} + 400\zeta(3) - 640\zeta(5)\right]L\right)\left(\frac{\alpha}{4\pi}\right)^4 \quad (3.1)$$

Here, the dependence on the external momentum $p$ is parametrized by the kinematic variable $L = \ln(-p^2/\mu^2)$ and $\alpha$ and $\xi$ respectively denote the $\widetilde{\text{MOM}}$ renormalized coupling and gauge parameter. Further it is reasonable to remark that our result reproduces the anomalous dimension previously derived in [14].

From this renormalized result, it transpires that the $\xi$ dependent terms in the self-energy up to the fourth order can be factorized

$$\Sigma(\alpha,\xi,L)|_{\alpha^4} = \Sigma(\alpha,0,L)\exp\left(-\xi L \frac{\alpha}{4\pi}\right)\bigg|_{\alpha^4} \tag{3.2}$$

yielding the expected exponential dependence where the minus in the exponent is due to the fact that we have considered the self-energy, which basically is the inverse of the connected electron propagator. Finally, we like to remark that the same pattern emerges in the $\overline{\text{MS}}$ scheme. However, it is more difficult to recognize due to the non-vanishing constant-$L$ terms in that scheme.

## 4. Diagrammatic characterization of the $\xi$ dependence

In the preceding sections, it transpired that the non-perturbative nature of the LKFT approach manifests itself in the fact that the gauge transformation only applies to the *external legs* of the electron propagator. However, in perturbation theory, high order Feynman graphs usually incorporate subgraphs with gauge dependent terms. As was demonstrated in [15], this discrepancy is resolved by systematically constructing and exploiting cancellations between the gauge dependent parts of dimensionally regularized Feynman graphs at $d = 4 - 2\varepsilon$ dimensions. This section contains our main result which is a simple diagrammatic expression that characterizes the $\xi$ dependence to all orders in perturbation theory.

First, we restrict ourselves to connected Green's functions with a single external fermion line whereas the number of external photon is arbitrary. Also, we will always amputate external propagators of photon legs. Thanks to this minor change, adding further external photon legs does not induce new terms in our diagrammatic formula which is contrary to the LKFT approach.

Then, we expand this Green's function as a series in the gauge parameter

$$G_c(\varepsilon,\alpha,\xi) = g_0(\varepsilon,\alpha) + g_1(\varepsilon,\alpha)\xi + g_2(\varepsilon,\alpha)\xi^2 + \ldots \tag{4.1}$$

and denote the coefficient of $\xi^n$ by $g_n$, which is a Laurent series in the dimensional regulator $\varepsilon$. The coefficient $g_0$ denotes the $\varepsilon$ expansion of $G_c$ in the Landau gauge and we shortly give a method to compute all higher coefficients $g_n$ with $n > 1$. Here, it should be remarked that the method allows to expand $G_c$ around other points than the origin $\xi = 0$ and hence can be equally applied to the Feynman ($\xi = 1$) or other specific gauges. For the sake of simplicity, we will expand around the Landau gauge in the following.

On the diagrammatic level, an expansion in the gauge parameter can be understood as a decomposition of the photon propagator into the Landau gauge part and a longitudinal part that depends on the gauge parameter – both parts are treated as separate propagators. If the number of longitudinal propagators is fixed to be $n$ and all connected diagrams are taken into account, then the coefficient $g_n$ is derived. Further, considering all possibilities to replace one propagator in the Landau gauge by a longitudinal propagator results in the coefficient $g_{n+1}$. As shown in [16], cancellations between diagrams of various topologies can be exploited to derive a closed diagrammatic expression that relates both coefficients

$$\boxed{g_{n+1}} = -\frac{1}{n+1} \boxed{g_n} . \qquad (4.2)$$

Here, we have introduced following set of auxiliary Feynman rules

$$= ie\delta_{ij} \qquad\qquad \cdots\bullet\cdots = i\xi \qquad (4.3)$$

$$i \xrightarrow{p} j = i\left(\frac{\not{p}-m}{\not{p}-m}\right)_{ij} = i\delta_{ij} \qquad \cdots\xrightarrow{p}\cdots = \frac{1}{p^2}. \qquad (4.4)$$

By iteration, equation (4.2) determines the higher-order coefficients in terms of the Landau gauge result $g_0(\varepsilon,\alpha)$ – it is a perturbative version of the LKFT naturally derived in momentum space. We have explicitly confirmed that an iteration of the epsilon expansion in the Feynman and the Landau gauge through equation (4.2) yields the correct gauge dependence up to fourth loop order.

Three remarks are in order.

1. Our characterization of the gauge dependence (4.2) implies that (after considering all cancellations) longitudinal photons effectively couple to external electron legs only. This explains why *internal photons are protected* from applying a gauge transform in the LKFT approach.

2. We like to emphasize that our approach holds for massless as well for massive fermions – the diagrammatics is completely the same, only the Feynman rule of the fermion propagator is adjusted for the fermionic mass *m*.

3. Our diagrammatic characterization immediately reveals that the on-shell limit does not depend on the gauge parameter $\xi$: application of the fermionic on-shell projectors to the right-hand side of (4.2) vanishes as the external propagators are amputated.

## 5. Higher dimensional QED

In this section, we report on a first application of the structural result (4.2) in a state of the art computation that has been conducted by Gracey [17] studying gauge theories in higher dimensions.

These enquiries are motivated due to the underlying universality that establishes relations between different gauge theories across different dimensions. In [17], Gracey constructed the follow-ing six and eight dimensional Lagrangians

$$\mathscr{L}_{d=6} = -\frac{1}{4}(\partial_\mu F_{\nu\rho})(\partial^\mu F^{\nu\rho}) - \frac{1}{2\xi}(\partial_\mu \partial^\nu A_\nu)(\partial^\mu \partial^\rho A_\rho) + i\overline{\psi}\slashed{\partial}\psi$$

$$\mathscr{L}_{d=8} = -\frac{1}{4}(\partial_\mu \partial_\nu F_{\rho\sigma})(\partial^\mu \partial^\nu F^{\rho\sigma}) - \frac{1}{2\xi}(\partial_\mu \partial_\nu \partial^\rho A_\rho)(\partial^\mu \partial^\nu \partial^\sigma A_\sigma) + i\overline{\psi}\slashed{\partial}\psi$$
$$+ \frac{g_2^2}{32}(F_{\mu\nu}F^{\mu\nu})(F_{\rho\sigma}F^{\rho\sigma}) + \frac{g_3^2}{8}F_{\mu\nu}F^{\nu\rho}F_{\rho\sigma}F^{\sigma\mu}$$

and showed that they lie in same universality class as the as four-dimensional Quantum Electrodynamics.

As the photon propagator in these higher dimensional field theories also decomposes into a physical transversal part and a gauge dependent longitudinal part, similar diagrammatic cancellations apply in these theories. In eight dimensions, quartic terms in the field-strength tensor yield quartic photon interactions. Of course, these terms potentially disturb the characterization of the gauge dependence (4.2). However, we proved that a contracted product of field-strength tensors always yields a transversal interaction in [16]. These transversal interactions decouple from the longitudinal part of the photon propagator. Therefore, the quartic photon vertices do not interfere in our derivation the gauge dependence and after a dimensional adjustment the characterization (4.2) also applies to the higher dimensional versions of Quantum Electrodynamics and reproduces the anomalous dimensions in $d = 6$ and $d = 8$ derived by Gracey.

## 6. The massless electron propagator

We observed that the gauge dependence in the renormalized electron propagator can be factorized into an exponential up to four loops. The characterization of the gauge dependence (4.2) explains this observation and generalizes to an arbitrary loop number. Furthermore, this proves that the anomalous dimension of the electron depends on the gauge parameter at the first loop order only.

First, we strip off a Lorentz factor from the electron propagator $\widetilde{S} = \slashed{q}S$. With this convention, the characterization of the gauge dependence (4.2) can be conveyed into momentum space

$$\frac{\partial \widetilde{S}}{\partial \xi}\left(\ln \frac{q^2}{\mu^2}, \xi\right) = ie^2 \int \frac{d^D p}{(2\pi)^D} \widetilde{S}\left(\ln \frac{(q+p)^2}{\mu^2}, \xi\right) \frac{\mathrm{Tr}\left[\slashed{q}(\slashed{q}+\slashed{p})\right]}{[p^2]^2 (q+p)^2}. \tag{6.1}$$

The factor $(p^2)^2$ in the momentum integration gives rise to an infrared divergence. In order to obtain a similar differential equation for the renormalized electron propagator, it is necessary to specify a boundary condition and subtract a counterterm which renders the infrared divergence finite and implements the boundary condition. The construction of such a counterterm for the $\widetilde{\mathrm{MOM}}$ scheme has been discussed in detail in [16]. Here, we only give the resulting differential equation for the renormalized massless electron propagator

$$\frac{\partial \widetilde{S}}{\partial \xi}\left(\ln \frac{-q^2}{\mu^2}, \xi\right) = \ln\left(\frac{-q^2}{\mu^2}\right) \frac{\alpha}{4\pi} \widetilde{S}\left(\ln \frac{-q^2}{\mu^2}, \xi\right), \tag{6.2}$$

which allows for the previously observed exponential solution

$$\widetilde{S}(L,\xi) = \widetilde{S}(L,0)\exp\left[\xi L \frac{\alpha}{4\pi}\right]. \tag{6.3}$$

Finally, we exploit the fact that both the electron propagator in the general covariant gauge and the Landau gauge have to fulfil the renormalization group equation. Further, recall that the renormalization group functions of the coupling and the gauge parameter are equal up to a sign $\beta = -\delta$. This implies that the anomalous dimension in the general covariant gauge and the Landau gauge just differ by the one-loop contribution

$$\gamma(\alpha,\xi) - \gamma(\alpha,0) = -\xi\frac{\alpha}{4\pi}. \tag{6.4}$$

This proves the statement that the anomalous dimension in $\widetilde{\mathrm{MOM}}$ scheme depends on the gauge parameter only at first loop order; see [16] for the $\overline{\mathrm{MS}}$ scheme.

## 7. Conclusion

For Quantum Electrodynamics in the linear covariant gauge, we verified cancellation between Feynman graphs and showed that the expansion in the gauge parameter of the sum over all connected graphs allows for a closed diagrammatic recurrence. This recurrence is described through a simple insertion into a one-loop graph and its iteration allows for the reconstruction of the general covariant gauge from a specific gauge (such as the Feynman or Landau gauge). We have explicitly checked that in the case of the massless electron propagator.

Based on this diagrammatic recurrence, we derived a differential equation for the renormalized massless electron propagator. Its solution features an exponential behaviour in the gauge parameter as observed in perturbative computations to four-loop order. Further, this proves the folklore statement that the anomalous dimension of the electron depends on the gauge parameter at first order in perturbation theory only.

It should be stressed that our approach is entirely based on the analysis of Feynman graphs – it is of purely perturbative nature. Also, we have been able to derive results well-known from the LKFT. Therefore, the characterization (4.2) of the gauge dependence can be considered as a perturbative version of the LKFT in momentum space.

It is an ongoing project to work out similar recurrences for non-abelian gauge theories. Due to the increased number of gauge boson interactions, one has to expect that a recurrence is not described by a single skeleton but by an infinite sum over such like skeletons.

The 't Hooft–Veltman gauge can be considered as a stepping stone towards a generalization of our results towards the non-abelian case. This is a non-linear gauge fixing for Quantum Electrodynamics and enhances the linear covariant gauge fixing by tri- and tetravalent photon interactions and a ghost sector. This non-linear gauge diagrammatically matches a non-abelian gauge theory but still allows for simple cancellations.

**Acknowledgments**

I am grateful to David Broadhurst for bringing the gauge dependence of the anomalous dimension of the electron and the folklore relating to that to my attention. Thanks to Dirk Kreimer for *On the gauge dependence of QED*    Henry Kißlerhis support and John Gracey for helpful comments on the manuscript. The use of axodraw2 [18] is acknowledged. This research was supported by Deutsche Forschungsgemeinschaft (DFG) through the grant KR 1401/5-1.## References

[1] L. D. Landau and I. M. Khalatnikov, *The gauge transformation of the Green function for charged particles*, Sov. Phys. JETP **2** (1956) 69.

[2] E. S. Fradkin, *Concerning some general relations of quantum electrodynamics*, Sov. Phys. JETP **2** (1956) 361.

[3] H. Sonoda, *On the gauge parameter dependence of QED*, Phys. Lett. **B499** (2001) 253 [hep-th/0008158].

[4] A. G. Grozin, *Introduction to effective field theories. 3. Bloch-Nordsieck effective theory, HQET*, [hep-ph/1305.4245].

[5] B. Zumino, *Gauge properties of propagators in quantum electrodynamics*, J. Math. Phys. **1** (1960) 1.

[6] P. Nogueira, *Automatic Feynman graph generation*, J. Comput. Phys. **105** (1993) 279.

[7] J. A. M. Vermaseren, New features of FORM. 2000 [math-phy/0010025].

[8] M. Tentyukov and J. A. M. Vermaseren, *The Multithreaded version of FORM*, Comput. Phys. Commun. **181** (2010) 1419 [hep-ph/0702279].

[9] S. G. Gorishnii, S. A. Larin, L. R. Surguladze and F. V. Tkachov, *Mincer: Program for Multiloop Calculations in Quantum Field Theory for the Schoonschip System*, Comput. Phys. Commun. **55** (1989) 381.

[10] S. A. Larin, F. V. Tkachov and J. A. M. Vermaseren, *The FORM version of MINCER*, NIKHEF-H-91-18 (1991) .

[11] B. Ruijl, T. Ueda and J. A. M. Vermaseren, *Forcer, a FORM program for the parametric reduction of four-loop massless propagator diagrams* [hep-ph/1704.06650].

[12] B. Ruijl, T. Ueda, J. A. M. Vermaseren and A. Vogt, *Four-loop QCD propagators and vertices with one vanishing external momentum*, JHEP **06** (2017) 040 [hep-ph/1703.08532].

[13] H. Kißler, *Hopf-algebraic Renormalization of QED in the linear covariant Gauge*, Annals Phys. **372** (2016) 159 [hep-th/1602.07003].

[14] D. J. Broadhurst, *Four loop Dyson–Schwinger–Johnson anatomy*, Phys. Lett. **B466** (1999) 319 [hep-ph/9909336].

[15] H. Kißler and D. Kreimer, *Diagrammatic Cancellations and the Gauge Dependence of QED*, Phys. Lett. **B764** (2017) 318 [hep-th/1607.05729].

[16] H. Kißler, *Computational and Diagrammatic Techniques for Perturbative Quantum Electrodynamics*. Ph.D. thesis, Humboldt-Universität zu Berlin, 2017.

[17] J. A. Gracey, *Six dimensional QCD at two loops*, Phys. Rev. **D93** (2016) 025025 [hep-th/1512.04443].

[18] J. C. Collins and J. A. M. Vermaseren, *Axodraw Version 2*. 2016 [cs.OH/1606.01177].7